\documentclass[reprint,aip,showpacs]{revtex4-1}

\usepackage{graphicx, dcolumn, bm}
\makeatletter

\newcommand{\Rmnum}[1]{\expandafter\@slowromancap\romannumeral #1@}
\usepackage[breaklinks=true,colorlinks,citecolor=blue,linkcolor=blue,urlcolor=blue]{hyperref}
\setcitestyle{super}
\makeatother

\begin{document}

\title{Transverse spin relaxation and magnetic correlation in $Pr_{1-x}Ca_{x}MnO_{3}$ : influence of particle size variation and chemical doping}

\author{Vinay Kumar Shukla}
\email{vkshukla@iitk.ac.in}

\author{Soumik Mukhopadhyay}
\email{soumikm@iitk.ac.in}
\affiliation{Department of Physics, Indian Institute of Technology, Kanpur 208 016, India}

%\date{\today}

\begin{abstract}
The short ranged magnetic correlations and dynamics of hole doped Pr$_{1-x}$Ca$_{x}$MnO$_{3}$ ($0.33<x<0.5$) of different crystallite sizes have been investigated using electron spin resonance spectroscopy (ESR). The major contribution to the temperature dependence of paramagnetic line-width is attributed to the spin-lattice relaxation dominated by thermally activated hopping of small polarons with typical activation energy of 20-50 meV. Irrespective of the crystallite size and dopant concentration, the transverse spin relaxation time ($t_2$) follows a universal scaling behaviour of the type $t_2\sim (T/T_0)^n$ in the paramagnetic regime, where $T_0$ and n are scaling parameters. Using the temperature dependence of $t_2$, we construct a phase diagram which shows that near half-doping, the magnetic correlations associated with charge ordering not just survives even down to the crystallite size of $22$ nm, but is actually enhanced. We conclude that the eventual suppression of charge ordering with reduction in particle size is possibly more to do with greater influence of chemical disorder than any intrinsic effect.

\end{abstract}


\maketitle

\section{Introduction}

The magnetic properties of mixed valence manganites are highly sensitive to the crystallite size variation as well as the variation in the dopant concentration~\cite{Zhang, Chai, Lu}.  Among these mixed valence manganites, hole doped Pr$_{1-x}$Ca$_x$MnO$_{3}$ ($0.33<x<0.5$)(PCMO) in bulk form shows transition from paramagnetic (PM) to antiferromagnetic (AFM) state below $T_N\sim140$ K and display the onset of charge ordering below $T_{CO}\sim240 K$. While there is considerable agreement that on changing the dopant concentration, such as if one moves away from half doping, the CO state is progressively weakened and suppressed~\cite{Liu}, there is less clarity regarding the influence of reduction of particle size on the charge ordered (CO) state. Despite several reports on the complete suppression of the CO state as well as the AFM ordering below a certain crystallite size~\cite{Zhang1, Pramanik, Rao, Jirak, Shukla1, Shukla2, Sarkar, Biswas}, the physical origin of the phenomenon remains unclear. Moreover, there are recent reports on persistence of short range spin-charge correlations and charge gap in nanocrystalline Nd$_{0.5}$Ca$_{0.5}$MnO$_{3}$~\cite{Zhou, Zhou1} and absence of significant correlation between size reduction and pressure effects on manganites in general~\cite{Shukla1, Giri}.

Electron spin resonance (ESR) spectroscopy is a preferred tool to probe the magnetic correlations and dynamics at microscopic level. While some ESR studies have focussed on the effect of variation of the dopant concentrations~\cite{Liu, Auslender1, Alejandro}, others investigated the influence of change in the crystallite size~\cite{Zhou, Shukla1, Shukla2, Auslender2}. The temperature dependence of ESR line-width in manganites is currently being debated and a consensus on the relaxation mechanism remains elusive~\cite{Seehra, Rozengerg1, Shames, Angappane1, Fan1, Fan2, Angappane2, Rozenberg2}. For example, in the paramagnetic regime, several mechanisms have been proposed to understand the temperature dependence of the ESR linewidth: 1) while studying the electron paramagnetic resonance spectrum of $(La, Y)_{2/3} (Ca, Sr, Ba)_{1/3}MnO_3$, Yuan \textit{et al.}~\cite{Yuan} suggested that the temperature dependence of line-width originates from the combination of exchange-narrowing spin-spin interactions and the spin lattice interactions; 2) Seehra \textit{et al.} proposed that the linear temperature dependence of linewidth in $La_{0.67}Sr_{0.33}MnO_3$ and $La_{0.62}Bi_{0.05}Ca_{0.33}MnO_3$ originates from spin-lattice interaction~\cite{Seehra}; 3) Causa \textit{et al.}~\cite{Causa} suggested a single relaxation mechanism, related to spin-only interactions, to explain the quasilinear dependence of linewidth; 4) Shengelaya \textit{et al.}~\cite{Shengelaya1,Shengelaya2} observed a striking similarity in the temperature dependence of paramagnetic linewidth and conductivity which was interpreted in terms of bottle-necked spin-lattice relaxation via conduction electrons.

In this article, we study the magnetic correlation and dynamics of $Pr_{1-x}Ca_xMnO_{3}$ ($0.33<x<0.5$) by changing the dopant concentration as well as the crystallite size from a few microns down to about $20$ nanometers using ESR spectroscopy. We demonstrate that while the CO state is suppressed by decreasing the dopant concentration from half doped (x=0.5) state, the effect of reduction of crystallite size on $T_{CO}$ is not the same.

\begin{figure*}
\includegraphics[width=12.5 cm]{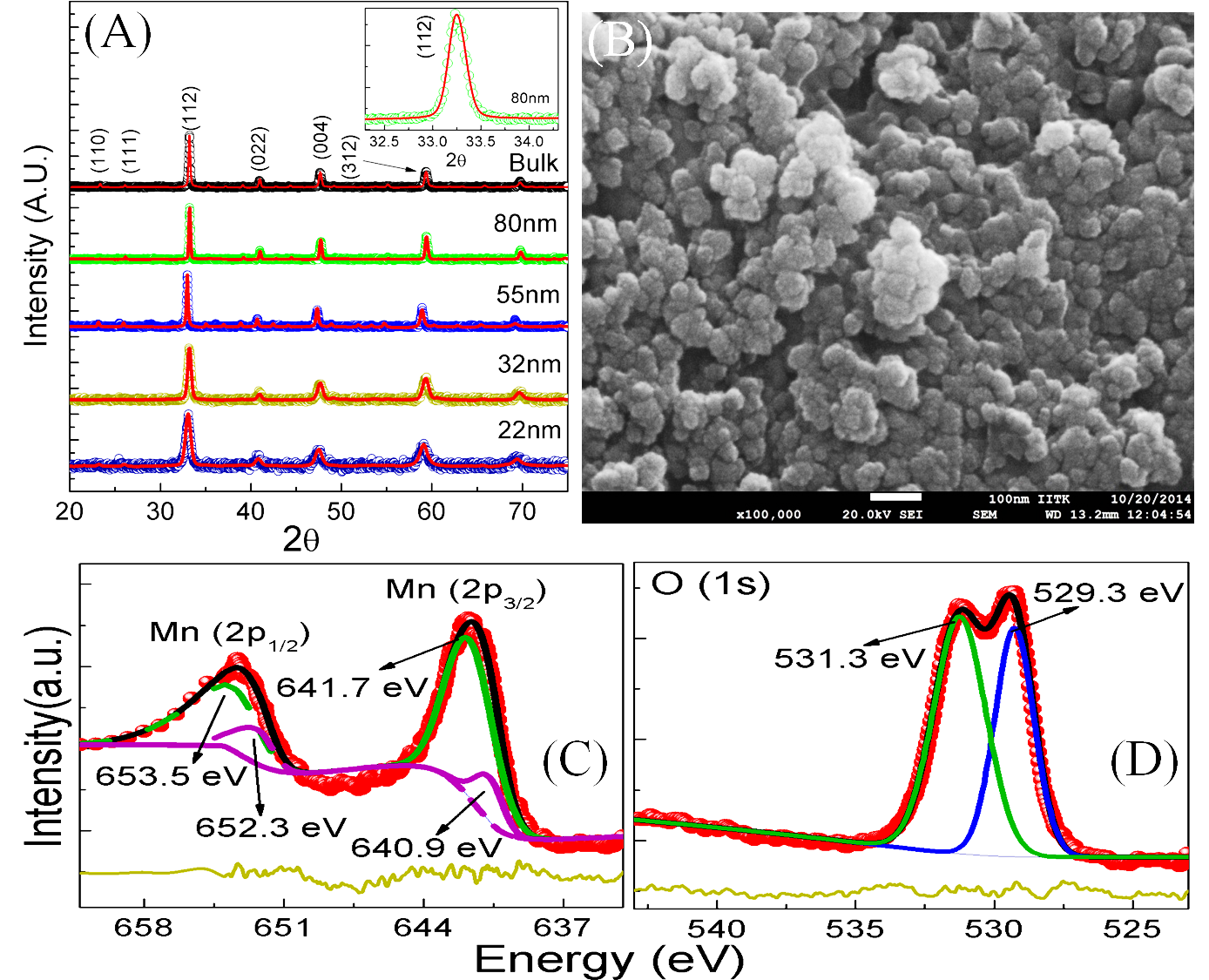}
\caption{(A) X-ray diffraction $\theta-2\theta$ scans for $Pr_{0.5}Ca_{0.5}MnO_{3}$ with different crystallite size. Solid red line shows the fit to the XRD patterns by Rietveld refinement. Inset: zoomed out of (112) peak for 80 nm sample. (B) FE-SEM image of PCMO (x=0.5, 22 nm). The average crystallite size is nearly $30$ nm, which is in good agreement with that calculated from Debye Secherrer formula. (C) and (D) X-ray photoemission spectra of Mn(2p) and O(1s), respectively. Solid black lines indicate the fits to XPS spectra. }\label{fig:xrd}
\end{figure*}

\section{Experiment, results and discussion}

Nanocrystalline $Pr_{1-x}Ca_xMnO_{3}$ (PCMO) (x=0.33, 0.45 and 0.5) were prepared by standard sol-gel synthesis route described elsewhere~\cite{Shukla1}. The structural characterization of all the samples were done by x-ray diffraction $\theta$- 2$\theta$ scans at room temperature using PANalytical X'pert diffractometer with $Cu-K_\alpha$ radiation having wavelength of 1.54 \AA. The microstructure, crystallite size and its distribution were studied by field emission scanning electron microscope (FE-SEM, Jeol, JSM-7100F). The x-ray photoemission spectroscopy was carried out using PHI 5000 Versa Prob II, FEI Inc. The magnetic measurements were done by PPMS (Quantum Design). The temperature dependent ESR spectroscopy measurements were done on powdered samples using Bruker EPR EMX spectrometer in X-band. The electrical transport properties were measured by Radiant Precession Premier II tester using I-V profile set up. Before the measurements, the powdered samples were pressed into pellets at a pressure of 40 kPa/$cm^2$ and contacts were made using silver paint.  

\subsection{Structural characterization}
Fig.~\ref{fig:xrd}(A) shows x-ray diffraction (XRD) pattern $\theta$- 2$\theta$ scans taken at room temperature for $Pr_{0.5}Ca_{0.5}MnO_{3}$ which reveals single phase nature of all the samples having different crystallite size. The Rietveld refinement analysis using full prof suite reveals that the room temperature phase of all the samples has orthorhombic structure having \textit{Pbnm} space group symmetry with lattice constants $\sim$ a= 5.41 $\AA$,  b=5.39 $\AA$, c= 7.61 $\AA$  and lattice volume of 221.9 $\AA^3$. The goodness of fit defined in terms of $\chi^2$ is below 1.5 for all the fits.  The crystallite size of all the nanocrystalline samples are estimated by Debye-Scherrer formula which are consistent with that obtained from FE-SEM images (Fig.~\ref{fig:xrd}B). The chemical composition of all the samples are confirmed by energy dispersive spectroscopy (EDS).

We have studied the electronic structure of all the samples by x-ray photoemission spectroscopy (XPS). The XPS data is fitted by the asymmetric Gauss-Lorentz sum function with Shirley background using XPS peakfit 4.1 software. Fig.\ref{fig:xrd}C shows a representative Mn(2p) XPS spectra for PCMO (x=0.5, 22 nm) which consists of two major peaks centred at 641.6 eV and 653.3 eV corresponding to Mn($2p_{3/2}$) and Mn($2p_{1/2}$) doublets. The peak centred at 641.6 eV (Mn($2p_{3/2}$)) is fitted with two peak components centred at 641.7 eV and 640.9 eV which are identified as contributions from $Mn^{4+}$ and $Mn^{3+}$ oxidation states, respectively. The ratio of areas covered  under the peaks centred at 641.7 eV and 640.9 eV gives the value 5.5 suggesting that the dominant contribution is coming from $Mn^{4+}$ oxidation state. Similarly, for Mn($2p_{1/2}$), two peaks centred at 653.5 eV and 652.3 eV are needed to get a proper fit, which suggests the contributions coming from $Mn^{4+}$ and $Mn^{3+}$ oxidation states with corresponding ratio of area-under-the-peaks being 3.8. On the other hand, Fig.\ref{fig:xrd}D shows the O(1s) XPS spectra for PCMO (x=0.5, 22 nm). Usually the O(1s) spectra consists of single peak centred at 529.3 eV, whereas from Fig.\ref{fig:xrd}D it is evident that two peaks centred at 531.3 eV and 529.3 eV are needed to fit the experimental data. This suggests presence of mixed oxidation states of Mn affecting the local environment of Mn-O-Mn bonds and leading to the formation of doublet in O(1s). Similar XPS spectra has been reported for other manganites~\cite{Lija, Kurmaev, Hui}.

\subsection{Magnetic characterization}

\begin{figure}
\includegraphics[width=8 cm]{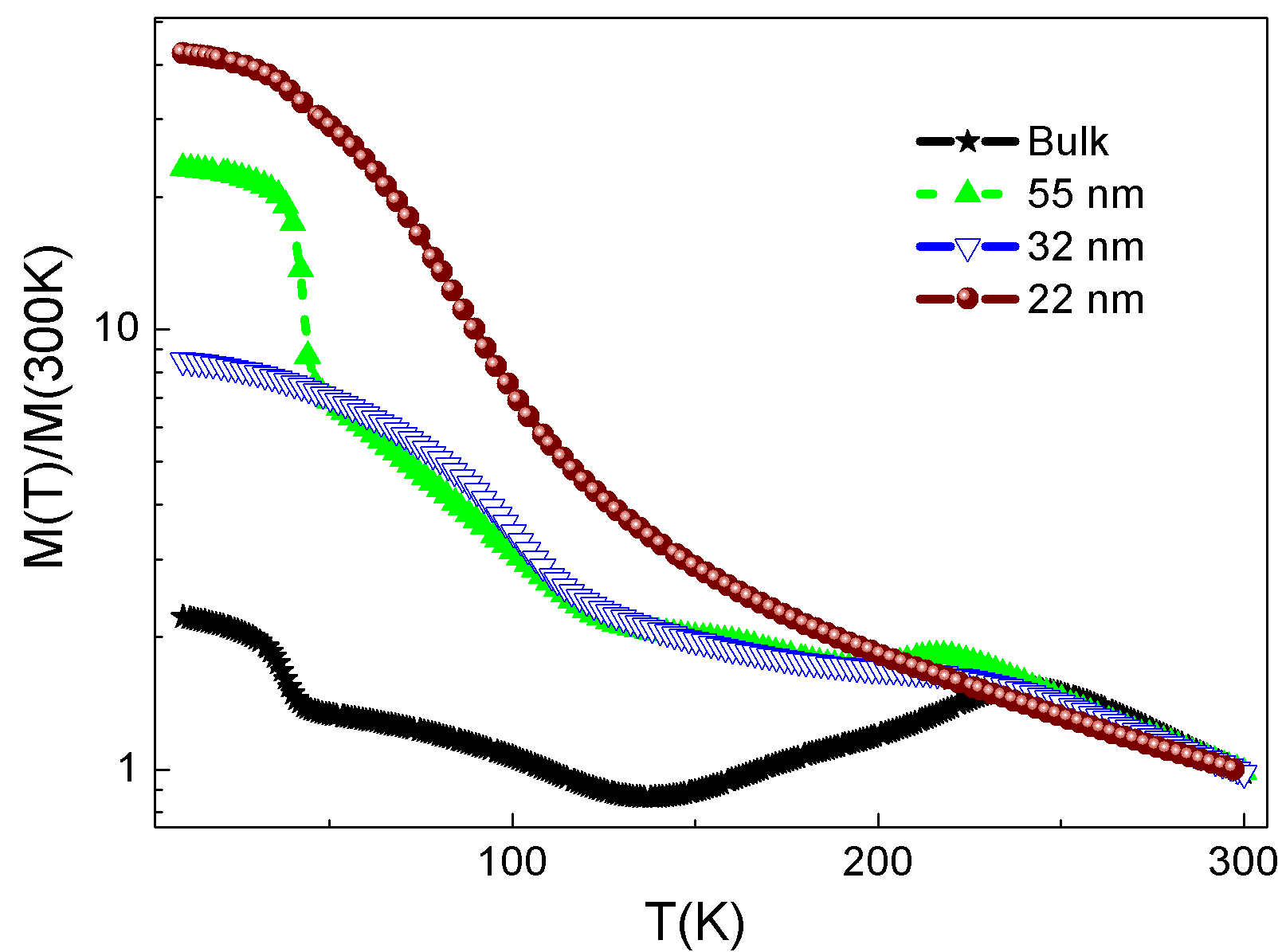}
\caption{Temperature dependence of field cooled (FC) dc magnetization normalized with respect to its room temperature value for bulk and nanocrystalline samples with different crystallite sizes for $Pr_{0.5}Ca_{0.5}MnO_{3}$. The applied magnetic field is 1 kOe.}\label{fig:susceptibility}
\end{figure}

The temperature dependence of field cooled dc magnetization is shown in Fig.~\ref{fig:susceptibility} for $Pr_{0.5}Ca_{0.5}MnO_{3}$ having different crystallite size. The anomaly around T$\sim$240 K is generally attributed to the onset of long range zener polaron (ZP) ordering with strong ferromagnetic correlation within $Mn^{3+}$ and $Mn^{4+}$ pairs. With the reduction of crystallite size, the anomaly associated with long range ZP ordering gets progressively broadened and eventually disappears completely below certain crystallite size. The samples in which ZP ordering gets completely suppressed are labeled as PCMO (x=0.5, 22nm), PCMO (x=0.45, 42 nm) and PCMO (x=0.33, 32 nm). For the polycrystalline (bulk) sample the anomaly at T$\sim$140 K represents the onset of antiferromagnetic (AFM) ordering. With the reduction of crystalline size the long range AFM ordering, too, is increasingly suppressed. The anomalies around T$\sim$30 K for the samples with larger crystallite size indicate the ordering of Pr moments at low temperature. The temperature dependence of inverse susceptibility for the samples with lowest crystallite size for different dopant concentrations are fitted with Curie-Weiss (CW) law, which give the following values of curie constant and curie temperature, respectively: $1)$  2.5 emu.K/mole, 69 K for PCMO (x=0.5, 22nm), $2)$ 1.5 emu.K/mole, 73 K for PCMO (x=0.45, 42 nm) and $3)$ 2.9 emu.K/mole, 74 K for PCMO (x=0.33, 32 nm). We have further estimated the effective magnetic moment for these samples from the temperature dependence of dc susceptibility using the relation $\mu_{eff} =(8 \chi_{dc} T)^{1/2}$ ~\cite{Blundell, Butera, Xu}. The high temperature limit of effective magnetic moment gives the value of 5.76 $\mu_B$ for  PCMO (x=0.5, 22nm), 7.13 $\mu_B$ for  PCMO(x=0.45, 42 nm and  6.36 $\mu_B$ for PCMO (x=0.33, 32 nm), respectively. The observed values are comparable to the contributions coming from all the $Mn^{3+}$ and $Mn^{4+}$ spins~\cite{Zakharov}. 

\subsection{Electron Spin Resonance Spectroscopy}

\begin{figure}
\includegraphics[width=8.5 cm]{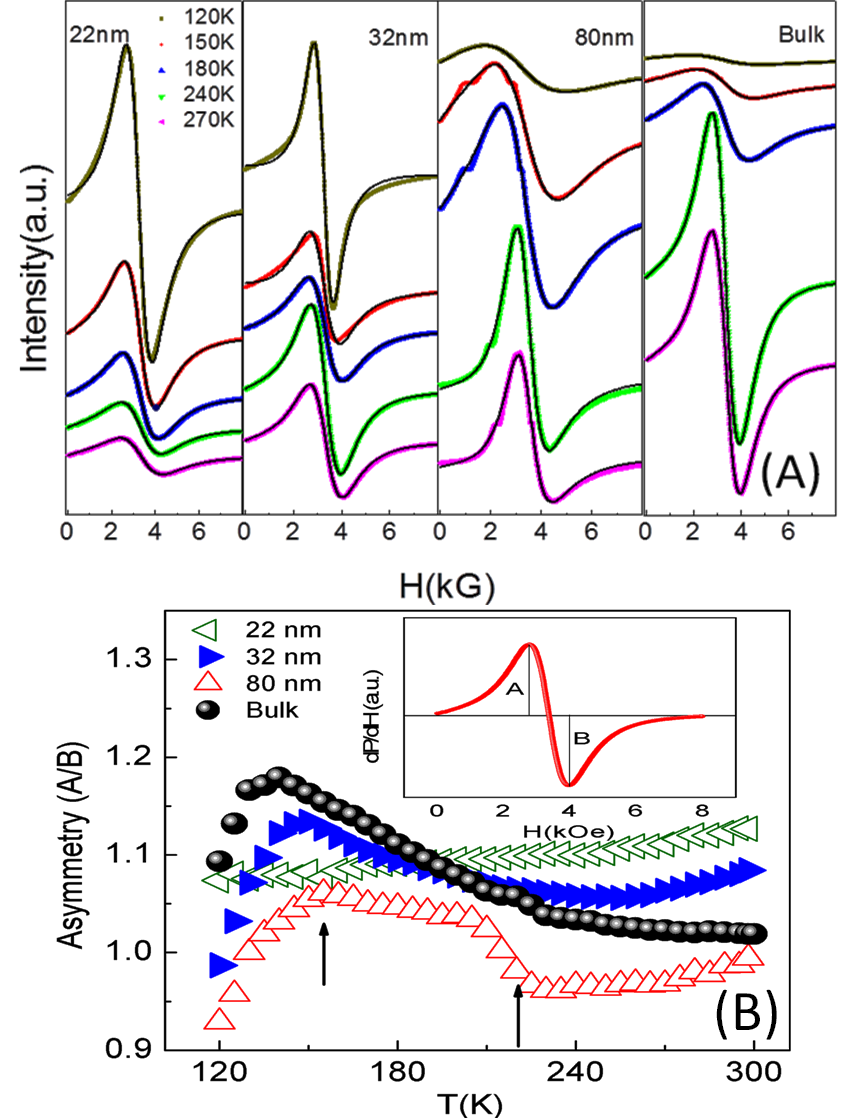}
\caption{(A) The differential ESR signals of $Pr_{0.5}Ca_{0.5}MnO_{3}$ for different crystallite sizes at some representative temperatures between 120K-300K with correponding fits. (B) Asymmetry factor (A/B) plotted against temperature for different crystallite sizes. Inset shows the low field maximum value A and high field minimum value B of ESR signals. The temperature dependence of A/B captures $T_N$ and $T_{CO}$ as indicated by arrow-heads.}\label{fig:esrsignal}
\end{figure}

The differential ESR signals were recorded at different temperatures  across $T_N$ and $T_{CO}$ leading to room temperature for all the samples. The samples were exposed to microwave radiation at constant frequency of 9.46 GHz (X-band) and external magnetic field was varied from 0 to 8000 Gauss. The power (P) absorbed by the sample from the transverse magnetic microwave field is captured in the form of its first derivative (dP/dH) by the standard lock-in technique. Fig.~\ref{fig:esrsignal}A shows the ESR signals of $Pr_{0.5}Ca_{0.5}MnO_{3}$ for different crystallite sizes at some representative temperatures between 120-300 K. In general, the lineshape is lorentzian. The temperature dependence of resonance field shows significant qualitative variations depending on doping and crystallite size. For example, in PCMO (x=0.5, 22 nm) the resonance field first decreases when cooled down from room temperature then shows a broad minima and decreases again on further cooling, whereas for PCMO (x=0.5, 32 nm) monotonic decrease in resonance field has been observed with decreasing the temperature (figure not shown).

The origin of ESR signals in these systems is attributed to the combined effect of $Mn^{3+}$ and $Mn^{4+}$ states (which are coupled through double exchange interaction) and the lattice~\cite{Lofland, Causa, Shengelaya1, Shengelaya2}. Fig.~\ref{fig:esrsignal}B represents the asymmetry factor (A/B) for x=0.5 samples, where A and B represents the extremum intensities at low and high field values respectively as shown in inset of ~\ref{fig:esrsignal}B. In case of powdered samples, since paramagnetic centres are randomly oriented, one could expect symmetric Lorentzian line shape. However, due to the internal fields associated with AFM/FM correlations and ZP ordering, asymmetry factor shows distinct anomalies as shown in Fig.~\ref{fig:esrsignal} which cannot be attributed to the artefact usually associated with single crystals~\cite{Kodera, Feher}. For others samples, too, we observe similar trends. For PCMO (x=0.5, 22 nm) in which ZP ordering is totally absent, we observe almost temperature independent (A/B) with value close to one.

\begin{figure}
\includegraphics[width=9 cm]{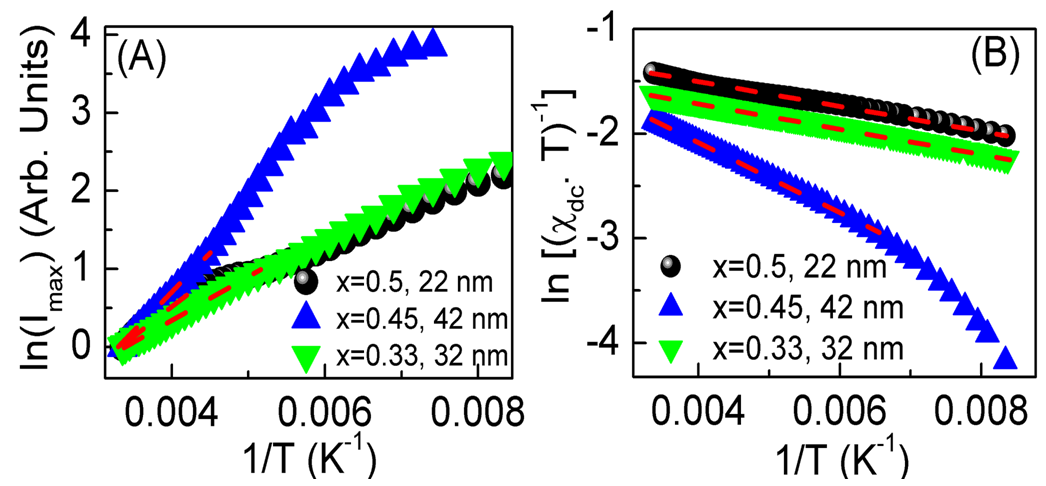}
\caption{Fitting of ESR spin susceptibilites (A) and dc susceptibilites  (B) by thermally activated models described in equation (1) and (2) respectively for different dopant concentrations (x).}\label{fig:thermalactivation}
\end{figure}

	In order to calculate the ESR intensity (I) we have integrated the differential ESR signals. In the paramagnetic regime the ESR intensity can be described by the thermally activated model~\cite{Yi, Oseroff, Causa}-
\begin{equation}
 I=I_{0} exp(E_{a1}/{K_{B}}T]
\end{equation} 	
where $I_0$ is a constant, $K_B$ is the Boltzmann constant and $E_{a1}$ is the activation energy. We have fitted ESR intensities using equation (1) as shown in Fig.~\ref{fig:thermalactivation}A. The activation energies ($E_{a1}$) thus obtained are given in Table I. In the paramagnetic regime, macroscopic magnetic susceptibility ($\chi_{dc}$) is directly proportional the ESR susceptibility ($\chi_{esr}$) with both $Mn^{3+}$ and $Mn^{4+}$ ions contributing to the overall ESR signal. The dc magnetic susceptibility can be fitted with the similar type of thermally activated model~\cite{Huber1} as
\begin{equation}
 1/\chi_{dc}= AT exp(-E_{a2}/{K_{B}}T]
\end{equation} 		
where A is a constant and $E_{a2}$ is the corresponding activation energy. Fig.~\ref{fig:thermalactivation}B shows the fitting of dc susceptibility data by equation (2). The fitting parameters are given in Table I. Clearly, for most of the samples, the activation energies obtained from the temperature dependence of ESR intensities are higher compared to that extracted from the temperature dependence of dc susceptibility. On the other hand, the high temperature limit of effective magnetic moment calculated from ESR susceptibility gives the value of 7.36 $\mu_B$ for  PCMO (x=0.5, 22nm), 7.15 $\mu_B$ for  PCMO (x=0.45, 42 nm) and  7.61 $\mu_B$ for PCMO (x=0.33, 32 nm) which are again larger compared to that calculated from dc susceptibility measurements. The enhancement of effective magnetic moment and activation energies calculated from ESR spin susceptibilities in comparison to that from dc susceptibilities has been reported before~\cite{Ivanshin, Causa} for $La_{1-x}Sr_{x}MnO_{3}$. Shengelaya \textit{et al.} interpreted this observation in terms of bottle-necked spin relaxation of exchange coupled $Mn^{3+}$/$Mn^{4+}$ spins to lattice via conduction electrons ~\cite{Shengelaya1,Shengelaya2}.
\subsection{Analysis of ESR linewidth}

\begin{figure}
\includegraphics[width=8 cm]{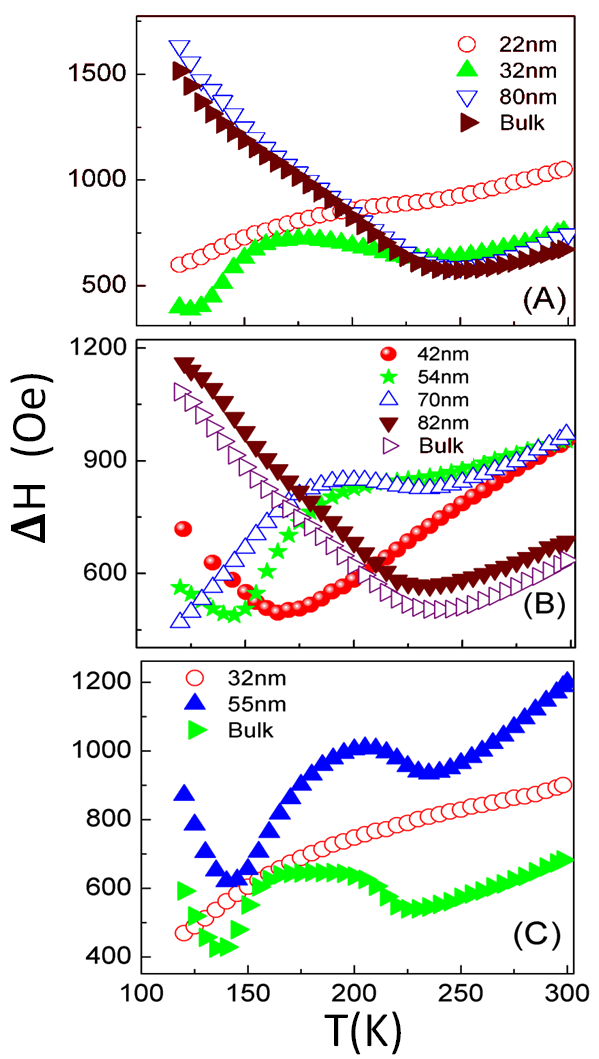}
\caption{Temperature dependence of ESR linewidth for different crystallite sizes of $Pr_{1-x}Ca_{x}MnO_{3}$: (A) x=0.5, (B) x=0.45, (C) x=0.33.}\label{fig:esrLW}
\end{figure}

We have fitted the differential ESR signals with lorentzian lineshape described in Ref.~\cite{Liu} for all the samples at different temperatures and obtained the linewidth and g factor as the fitting parameters. Fig.~\ref{fig:esrLW}A shows the temperature dependence of ESR linewidth ($\Delta H$) for $Pr_{0.5}Ca_{0.5}MnO_{3}$ corresponding to different crystallite sizes. For the bulk PCMO (x=0.5, bulk), $\Delta H$ first decreases with increasing temperature till $T_{CO}$, passes through a minima around $T_{CO}$ and then increases again with increasing temperature above $T_{CO}$. The increase in linewidth above $T_{CO}$ can be interpreted as an opening of the ``bottleneck" related to spin-lattice relaxation with increasing temperature as predicted by Shengelaya \textit{et al.} ~\cite{Shengelaya1, Shengelaya2}. Similar observations are made for the nanocrystalline PCMO (x=0.5, 80 nm). In contrast, the nanocrystalline PCMO (x=0.5, 32nm) shows an additional shallow minimum around 120 K. The change in sloped or increase in linewidth below 135 K in PCMO (x=0.5, bulk, 80nm, 32nm) indicates the onset of anti-ferromagnetic ordering (see Fig.~\ref{fig:susceptibility} for comparison). Similar observations are noted for other samples too as shown in Fig.~\ref{fig:esrLW}B and C. On further reducing the crystal size to 22 nm as in case of PCMO (x= 0.5, 22 nm), the linewidth decreases monotonically with decreasing temperature with a shallow minima around 240 K suggesting short range spin-charge correlations. This is interesting as the temperature dependence of field cooled dc magnetization indicates the absence of any long range charge ordering for PCMO (x= 0.5, 22 nm). On the other hand, as one moves away from half doping, even the short range correlations seems to get suppressed below a certain crystallite size as shown for PCMO (x=0.45, 42nm) and PCMO (x=0.33, 32nm).

\begin{figure}
\includegraphics[width=9 cm]{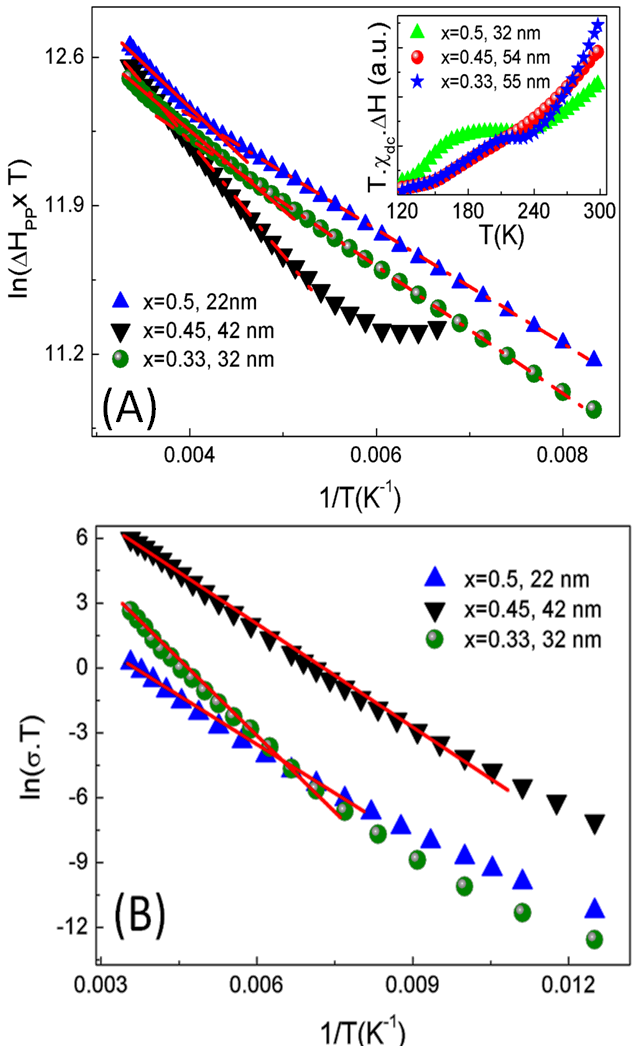}
\caption{(A) Fitting of ESR linewidth by Shengelaya \textit{et al.} ~\cite{Shengelaya1, Shengelaya2} model for some representative samples. (B) DC conductivity fitted with adiabatic small polaron hopping model is shown for comparision with ESR linewidth.}\label{fig:fitting}
\end{figure}

	In order to understand, the relaxation mechanism associated with ESR linewidth, two equivalent approaches have been proposed in the literature. First model is based on general theories of spin relaxation described by Huber \textit{et al.}~\cite{Huber} which suggests that away from the critical regions of  magnetic and structural phase transitions,  the temperature dependence of ESR linewidth can be related to dc susceptibility~\cite{Huber, Tovar, Causa, Deisenhofer, Atsarkin1} as follows.
\begin{equation}
 \Delta H = \frac{C\Delta H(\infty )}{T \chi_{dc}}
\end{equation}
where $C$ is Curie constant and temperature independent parameter $\Delta$H($\infty $) is defined as the high temperature limit of ESR linewidth. The values of the parameters obtained from fitting the experimental data with equation(3) are as follows:  $C= 2.9$ emu.K/mole.Oe and $\Delta H(\infty)= 1191$ Oe for PCMO (x=0.5, 22 nm); $C= 2.2$ emu.K/mole.Oe and $\Delta H(\infty)= 1296$ Oe for PCMO (x=0.45, 42 nm); and $C= 3.3$ emu.K/mole.Oe and $\Delta H(\infty)= 951$ Oe for PCMO (x=0.33, 32 nm). The second model prescribed by Shengelaya \textit{et al.}~\cite{Shengelaya1, Shengelaya2} suggests that the relaxation of exchange coupled localized spins ($Mn^{4+}:{t_{2g}}^3{e_{g}}^0$) to the lattice occurs via conduction electrons ($Mn^{3+}:{t_{2g}}^3{e_{g}}^1$), with the temperature dependence of the line-width following thermally activated behaviour-	
\begin{equation}
 \Delta H = \Delta H_{0} + {\frac{B}{T}}\exp\left(\frac{-E_{a3}}{K_{B}T}\right)
\end{equation}
where $B$, $\Delta H_{0}$ are constants and $E_{a3}$ is the activation energy for small polaron hopping. We have calculated the activation energies ($E_{a3}$) from fitting the temperature dependence of linewidth with equation(4) as shown in Fig.~\ref{fig:fitting}A. In the adiabatic limit, the temperature dependence of dc electrical conductivity in the small polaron picture can be expressed as $\sigma=(\sigma_0/T)exp(-E_c/kT)$, where $E_c$ is the activation energy and $\sigma_0$ is a constant. Fig.~\ref{fig:fitting}B describes the fitting of conductivity data by small polaron model and corresponding activation energies ($E_{c}$) are calculated from the slope of the linear portion in the semi-log plot. Table I encapsulates the comparison of activation energies $E_{a1}$, $E_{a2}$ and $E_{a3}$ obtained from ESR susceptibility, dc susceptibility and ESR linewidth, respectively, with the activation energy ($E_{c}$) obtained from dc conductivity in the paramagnetic regime.

	Fig.~\ref{fig:fitting} shows that for nanocrystalline samples which do not show long range charge ordering, the ESR linewidth follows similar temperature dependence as that of dc electrical conductivity indicating single phonon spin lattice relaxation via conduction electrons, the so called `bottleneck' regime. The absence of any spin-lattice contributions to ESR linewidth would anyway lead to temperature independent behaviour of the product $\Delta H T \chi_{dc}$ in the paramagnetic regime~\cite{Huber2, Yang2}, which is clearly not the case here (Inset, Fig.~\ref{fig:fitting}A). The activation energies obtained from linewidth are smaller when compared to that obtained from the corresponding temperature dependence of conductivity. Similar discrepancies between $E_{a3}$ and $E_{c}$ have been reported for other hole doped manganites~\cite{Yang2, Markovich, Misra} which are interpreted as follows: in manganites, there are two systems of thermally activated electrons (i.e. $t_{2g}$ and $e_g$) whose spins are coupled together by Hund's correlations. The magnetic properties are determined by $t_{2g}$ core spins whereas the electrical conductivity originates from the adiabatic hopping of $e_g$ small polarons. The hopping weakens the Hund's rule correlation between $t_{2g}$ core spins and the spins of $e_g$ polarons, which eventually gets suppressed with increasing $Mn^{3+}$-O-$Mn^{4+}$ distance or decreasing the $Mn^{3+}$-O-$Mn^{4+}$ bond angle, leading to considerable differences between $E_{a3}$ and $E_{c}$ values.
	
\begin{figure}
\includegraphics[width=8.8 cm]{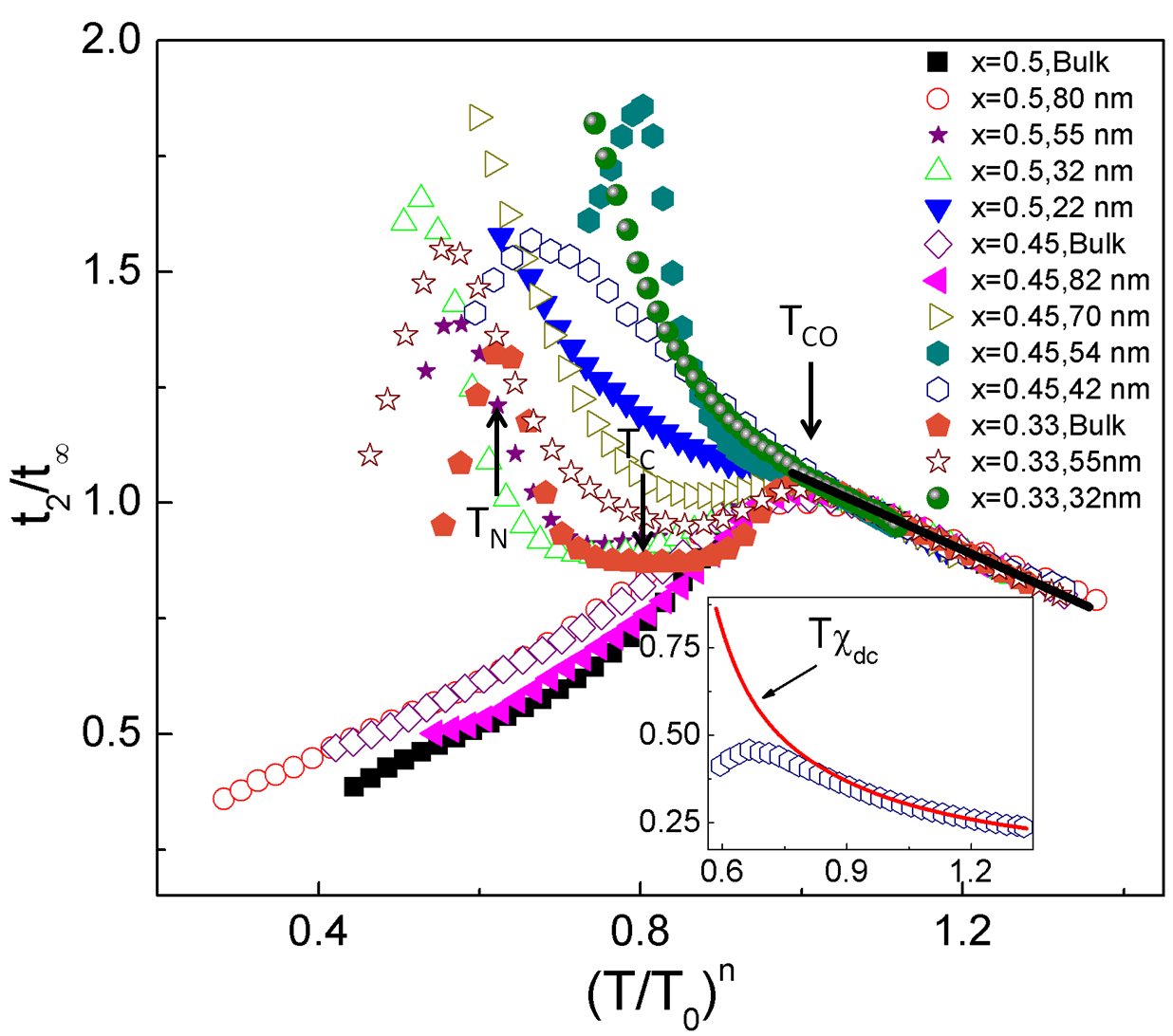}
\caption{Scaling of the transverse relaxation time ($t_2$) in the paramagnetic regime for all the samples of PCMO with different crystallite sizes and dopant concentrations. Inset: comparision of $t_2$ with product of dc magnetic susceptibility and temperature ($T\chi_{dc}$) as per non-critical Huber's Law~\cite{Huber} indicated by solid red line shown for a representative sample PCMO(x=0.45, 42 nm). The arrow-heads indicate $T_{CO}$, $T_N$ and $T_C$ for a representative curve (x=0.33, Bulk).}\label{fig:scaling}
\end{figure}

	We have studied the spin dynamics by calculating  the transverse relaxation time $t_2$ (also known as spin-spin relaxation time) from the ESR linewidth ($\Delta H$) using the following relation
\begin{equation}
(1/t_2)^{-1} = (\sqrt{3}/2) \gamma \Delta H
\end{equation}
where $\gamma$ is the electronic gyromagnetic ratio. It is interesting to note that transverse relaxation time follows a common scaling behaviour $t_2\sim (T/T_0)^n$ in the paramagnetic regime irrespective of the dopant concentration and crystallite size (Fig.~\ref{fig:scaling}).

\begin{table}[h!]
\centering
\caption{Energy scales and scaling parameters}
\begin{tabular}{|c|c|c|c|c|c|c|c|}
 \hline
 Sample&$E_{a1}$&$E_{a2}$&$E_{a3}$&$E_{c}$&$t_\infty$&n&$T_0$\\
 %\hline
 		&(meV)&(meV)&(meV)&(meV)&(ns)& &(K)\\
 \hline
 $x=0.50, Bulk$ & 45.6 & 59.8 & 48.6 & 99.4 & 0.82 & 1.11 & 246 \\
 \hline
 $x=0.50, 80 nm$ & 46.4 & 52.8 & 55.7 & 95.8 & 0.81 & 1.73 & 246 \\
 \hline
 $x=0.50, 55 nm$ & 58.3 & 44.7 & 53.2 & 97.1 & 0.46 & 0.99 & 225 \\
 \hline
 $x=0.50, 32 nm$ &  28.4 & 41.8 & 45.4 & 99.2 & 0.74 & 1.01 & 230 \\
 \hline
 $x=0.50, 22 nm$ &  45.2 & 13.2 & 23.2 & 104.2 & 0.50 & 0.69 & 220 \\
 \hline
 $x=0.45, Bulk$ & 132.7 & 29.6 & 53.2 & 194.7 & 0.95 & 1.21 & 238 \\
 \hline
 $x=0.45, 80 nm$ & 67.3 & - & 43.2 & 136.2 & 0.84 & 0.91 & 230 \\
 \hline
 $x=0.45, 70 nm$ & 63.1 & - & 36.7 & 140.1 & 0.57 & 0.75 & 233 \\
 \hline
 $x=0.45, 54 nm$ & 63.7 & 12.3 & 32.8 & 197.3 & 0.54 & 0.45 & 225 \\
 \hline
 $x=0.45, 42 nm$ & 92.1 & 28.2 & 44.1 & 137.4 & 0.63 & 1.16 & 225 \\
 \hline
 $x=0.33, Bulk$ & 202.7 & 31.8 & 44.3 & 198.2 & 0.80 & 0.91 & 230 \\
 \hline
 $x=0.33, 55 nm$ & 148.1 & 35.7 & 49.5 & 196.7 & 0.79 & 1.14 & 235 \\
 \hline
 $x=0.33, 32 nm$ & 47.4 & 10.2 & 26.2 & 188.3 & 0.71 & 0.45 & 236 \\
 \hline
\end{tabular}
\end{table}
The scaling parameters are given in Table I. When we compare the temperature dependence of transverse relaxation time $t_2$ with the product of dc magnetic susceptibility and temperature ($T\chi_{dc}$) as shown in inset of Fig.~\ref{fig:scaling}, it is evident that the observed scaling behaviour is consistent with the Huber's formula~\cite{Huber} for the non-critical temperature (paramagnetic) range: $(t_2)^{-1}$= Constant/($T\chi_{dc}$). Moreover, away from the magnetic and structural transitions the spin-spin relaxation time ($t_2$) is equivalent to spin-lattice relaxation time ($t_1$) leading to the universal scaling behaviour in the paramagnetic regime. The transverse relaxation time $t_2$ typically ranges from $1.1 \times 10^{-9}$ to $3.1 \times 10^{-10}$s which is consistent with the previous reports on other manganite systems ~\cite{Atsarkin2, Simon, Atsarkin1}.

	On the other hand, the increase in relaxation time ($t_2$) with decreasing temperature in the low temperature regime observed for most of the nanocrystalline samples (except for those with large crystallite size) can be interpreted as follows. The transverse relaxation time $t_2$ can be expressed as $(1/t_2)\sim(\gamma H_i)^2 \tau$, where $(H_i)^2$ is the mean-square amplitude of the fluctuating internal field, and $\tau$ is the correlation time. At a fixed correlation time, $(H_i)^2$  decreases with cooling due to onset of ferromagnetic correlations between Mn-Mn spins, leading to ``slowing down" of transverse spin relaxation. Conversely, the prevalence of anti-ferromagnetic correlation (in polycrystalline samples and those having larger crystallite size) leads to the enhancement of $(H_i)^2$ which ``speeds up" the relaxation phenomena. Similar observations regarding the low temperature behaviour have been reported for $La_{1-x}Ca_{x}MnO_3$ and $La_{2-2x}Sr_{1+2x}Mn_2O_7$~\cite{Simon, Atsarkin1}. Consequently, the transition temperatures associated with AFM($T_N$), FM ($T_C$) and charge ordering ($T_{CO}$) can be identified as shown in Fig.~\ref{fig:scaling}.

\begin{figure}
\includegraphics[width=8.5 cm]{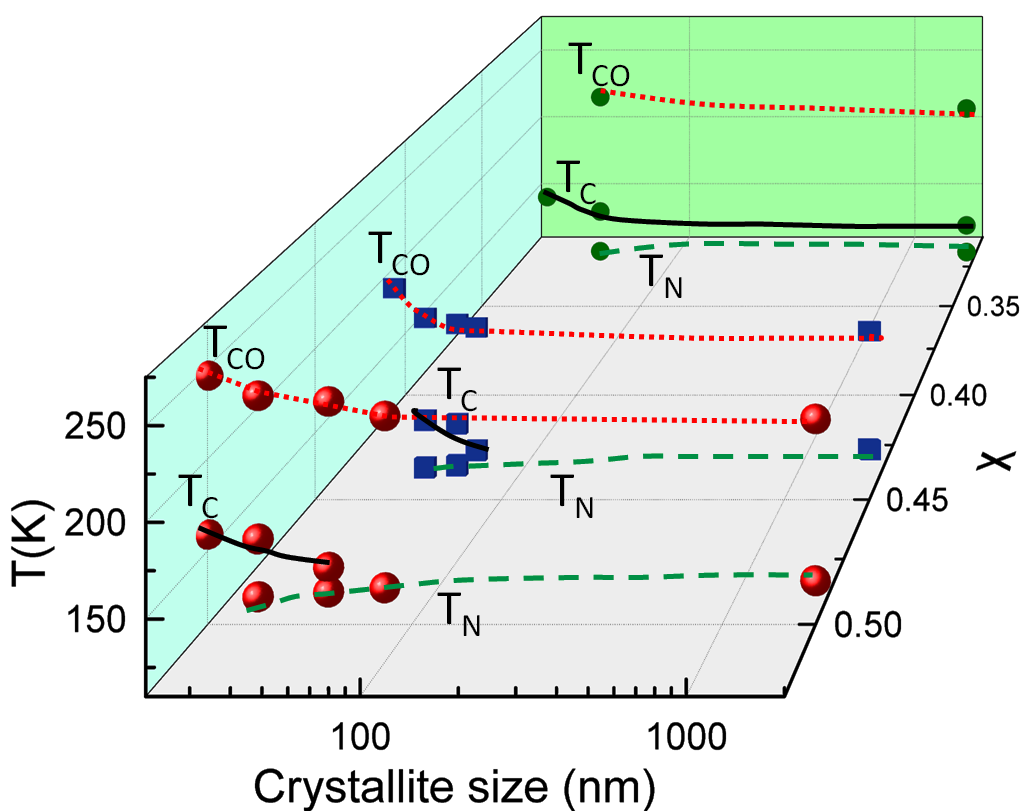}
\caption{Phase diagram constructed from the temperature dependence of $t_2$ for all the samples with different crystallite sizes and dopant concentrations. The values of $T_{CO}$, $T_N$ and $T_C$ of PCMO at different dopant concentrations (x) is plotted on the particle size- temperature plane.  Dashed lines correspond to the variation of $T_N$, dotted lines describes the variation of $T_{CO}$ and continous line represent the variation of $T_C$ with particle size. The three symbols correspond to three different values of x.}\label{fig:phasedia}
\end{figure}

	Using the temperature dependence of $t_2$ in Fig.~\ref{fig:scaling}, we construct a phase diagram that indicates the transition temperatures $T_N$, $T_C$ and $T_{CO}$ in Pr$_{1-x}$Ca$_x$MnO$_3$ ($0.3<x<0.5$) for different crystallite size and dopant concentrations (Fig.~\ref{fig:phasedia}). The values of $T_{CO}$, $T_C$ and $T_N$ have been calculated from the derivative plots of the temperature dependence of $t_2$ for all the samples for better accuracy. Curiously, with decreasing the crystallite size, $T_{CO}$ shows monotonic increase for the half dopant concentration. Similar observations were recorded for other samples, too. However, as we move towards x=0.3, the charge ordering gets completely suppressed below a certain crystallite size. Consequently, there is no $T_{CO}$ for PCMO (x=0.45, 42 nm) and PCMO (x=0.33, 32 nm). It is interesting to note that the upward shift of $T_{CO}$ with reduction in crystallite size calculated from microscopic transverse relaxation time in the present case shows good agreement with our previous report of the same estimated from macroscopic dc magnetization data~\cite{Shukla1}. This observation deserves closer scrutiny as the weakening of charge ordering and stabilization of ferromagnetic phase with reducing the crystallite size is generally attributed to the enhancement of surface pressure on the crystal structure and an increasing $e_g$ electron bandwidth with enhanced ferromagnetic interaction. According to the charge and orbital ordering picture, ferromagnetic correlations arising from the disordered and dynamic double exchange (DE) is quenched on approaching the $T_{CO}$ from above with the onset of charge ordering favouring the antiferromagnetic correlations leading to the magnetic anamoly at $T_{CO}$~\cite{Bao, Millange}. In contrast, the Zener Polaron ordering picture suggests that the magnetic anomaly observed at $T_{CO}$ is associated with formation of `units' of ferromagnetic pairs and `switching on' of antiferromagnetic (AFM) correlations between these `units' rather than suppression of DE, while the system as a whole remains in the paramagnetic state~\cite{Aladine}. It is clear from Fig.~\ref{fig:phasedia} that $T_N$ undergoes a weak downward shift with reduction in particle size. Now if the switching on of AFM correlations is only weakly affected by reduction in particle size, the magnetic anomaly associated with $T_{CO}$ should shift towards higher temperature with increased FM correlation between neighbouring pairs within the `unit'. This automatically suggests that eventual suppression of charge ordering with reduction in particle size as is commonly observed can only happen due to increased chemical disorder effect~\cite{Chen} with high surface to volume ratio in nanoparticles and not due to any other `intrinsic' cause.

\section{Conclusions}

The study of spin dynamics using ESR spectroscopy in $Pr_{1-x}Ca_{x}MnO_{3}$ for different dopant concentration of x=0.5, 0.45 and 0.33 and varying crystallite size suggests the inequivalent nature of the underlying physics at a fundamental level between chemical doping and reduction of particle size in these systems. The transverse spin relaxation time shows a universal scaling behaviour in the paramagnetic regime irrespective of dopant concentration and crystallite size. The phase diagram constructed out of the temperature dependence of $t_2$ at lower temperature clearly shows enhancement of the magnetic correlation associated with $T_{CO}$, rather than its suppression with reduction in particle size.

\section{Acknowledgements}
VKS acknowledges CSIR, India for the financial support.

\end{document}